\documentclass{article}

\usepackage{PRIMEarxiv}

\usepackage[utf8]{inputenc} 
\usepackage[T1]{fontenc}    
\usepackage{hyperref}       
\usepackage{url}            
\usepackage{booktabs}       
\usepackage{multirow}
\usepackage{amsmath,amssymb,amsfonts}%
\usepackage{amsthm}%
\usepackage{mathrsfs}%
\usepackage{nicefrac}       
\usepackage{microtype}      
\usepackage{lipsum}
\usepackage{fancyhdr}       
\usepackage{graphicx}       
\graphicspath{{media/}}     
\usepackage[finalnew]{trackchanges} 

\title{Robust data driven discovery of a seismic wave equation
}

\author{
  Shijun Cheng and Tariq Alkhalifah\\
  King Abdullah University of Science and Technology \\
  Thuwal 23955-6900\\
  \texttt{{sjcheng.academic@gmail.com}, tariq.alkhalifah@kaust.edu.sa} \\
}

\begin{document}
\maketitle

\begin{abstract}
\change{In spite}{Despite the fact} that our physical observations can often be described by derived physical laws, \change{like}{such as} the wave equation, in many cases, we observe data that do not match the laws or have not been described physically yet. \change{So}{Therefore} recently, a branch of machine learning \change{have}{has} been devoted to the discovery of physical laws from data. We test such discovery algorithms, with our own flavor of implementation \add{D-WE}, in discovering the wave equation \remove{D-WE} from the observed spatial-temporal wavefields. D-WE first pretrains a neural network (NN) in a supervised fashion to establish the mapping between the spatial-temporal locations $\left (x,y,z,t \right)$ and the observation displacement wavefield function $u\left (x,y,z,t \right)$. The trained NN serves to generate meta-data and provide the time and spatial derivatives of the wavefield (e.g., $u_{tt}$ and $u_{xx}$) by automatic differentiation. Then, a preliminary library\change{, including countable candidate function terms,}{of potential terms for the wave equation} is optimized from an overcomplete library by using \add{a} genetic algorithm. \change{Finally, we}{We, then,} use a physics-informed information criterion to evaluate the precision and parsimony of potential equations in the preliminary library and determine the best structure of \change{a}{the} wave equation\change{, and}{. Finally,} we \remove{further}train \change{a}{the} \add{"}physics-informed\add{"} neural network to identify the corresponding coefficients of each functional term. Examples in discovering the 2D acoustic wave equation validate the feasibility and effectiveness of D-WE. \change{Also,}{We also verify} the robustness of this method by testing it on noisy and sparse\add{ly acquired} wavefield data.
\end{abstract}

\section{introduction}
Seismic wave equations play a critical role in seismology to constrain and define wave propagation in a specific medium. Having an accurate equation contributes to our understanding of wave propagation. \change{Formerly, s}{S}eismologists \change{follow}{previously followed} some basic physical laws to propose numerous wave equations \cite{kjartansson1979constant, thomsen1986weak, alkhalifah2000acoustic, zhu2014modeling, cheng2021wave}. The presented equations provide the central ingredient to describe and model seismic wave propagation in the Earth’s interior. However, we can not ignore that many \add{these} equations are derived based on some approximations and assumptions. Moreover, wave propagation in \change{some}{certain} media may involve elusive and complex mechanisms \cite{biot1955theory, dvorkin1993dynamic, ba2008mesoscopic, ba2017rock}. Hence, the resulting governing equations might not accurately describe the wavefield, \change{like}{such as} in attenuating media, \change{which}{as they inherently} depend on the underlying assumptions \cite{hao2021nearly}. In this case, a natural solution is to seek a new accurate mathematical equation based on existing knowledge and physical principles to replace the original wave equation. Actually, it’s undeniable that this process can be difficult and time consuming. Then, we ask ourselves: can we discover a sufficiently precise wave equation directly from \change{observation}{observed} data without relying on physical laws?

Benefiting from the recent advances in machine learning (ML) and data-processing capabilities, the dawn of this question maybe in the horizon. More recently, data-driven discovery methods \change{are}{have been} developed to identify the underlying partial differential equations (PDEs) of physical problems. \change{Commonly}{Specifically}, data-driven discovery methods \change{need to}{are based on} construct\add{ing} a library composed of candidate function\add{al} terms, and then \change{use}{using} various optimization algorithms to select the most appropriate combination of candidate \change{sets}{terms}, generating the general form of \add{the} equations. In terms of library construction approaches, current methods are mainly divided into closed and expandable library methods.

Closed library methods, which are the most widely \change{applied}{used}, first build an overcomplete library and then use sparse regression methods to extract the dominating candidate function terms. \remove{In which, }Lasso \cite{schaeffer2017learning}, sequential threshold ridge regression \cite{rudy2017data}, and SINDy \cite{brunton2016discovering} are the leading representative sparse regression methods for discovering PDEs from the \add{closed} construction library. Although sparse regression methods remain highly efficient, the fact is that they are limited to \add{a} complete candidate library given beforehand. It’s difficult to provide an overcomplete candidate library that must contain true PDE terms, especially in a setting in which we lack prior knowledge. The growing size of candidate libraries, meanwhile, leads to the increasing difficulty of sparsifying them, which means that we may discover a wrong PDE. Furthermore, in the process of constructing a candidate library, the derivatives of the observation data are often obtained by using finite difference, polynomial interpolation and other methods. This is not robust in the face of data acquired on irregular grids and maybe exposed to noise.

In contrast, expandable library methods have a stronger ability to identify PDEs with complicated structures than the closed library methods. This is because expandable library methods don’t require a pre-determined overcomplete library. \change{It}{They} just require\remove{s} a randomly generated incomplete initial library, which will evolve to produce unlimited combinations \change{via}{through} the introduction of cross-over and mutation operations of genetic algorithms (GA) \cite{maslyaev2019data}.  However, processing noisy and sparse data still remains a challenge. With the rapid development of machine learning (ML), some researchers \cite{chen2021physics,xu2022discovery} utilized NN \add{functional representation} to calculate derivatives through automatic differentiation. Compared \change{with}{to} conventional numerical methods, the calculation of derivatives provided by NN is more stable and robust to noisy data \cite{xu2022discovery}. Meanwhile, NNs can operate in mesh free environments when computing derivatives via automatic differentiation. However, automatic differentiation is relatively costly, but for discovering an equation, \add{we believe} the accuracy is worth the cost.

In this work, we adapt a new data-driven method to discover the wave equation, named D-WE, which combines a neural network (NN) and GA. In D-WE, a fully connected deep NN is first trained, where the input to the network is spatial-temporal locations (e.g., $x,y,z,t$), and the corresponding output is \change{observation}{the observed} pressure wavefield $u\left(x,y,z,t \right)$. After training the network, we can produce meta-data and compute time and spatial derivatives of $u\left(x,y,z,t \right)$. Subsequently, a digital coding is created to define the form of the underlying wave equation, including the left-hand side (LHS) and right-hand side (RHS). The special encoding corresponds to the genomes for the combination of some candidate function terms. Cross-over and mutation are employed to expand the diversity of the library, that is, increasing the search scope of candidate wave equations. To determine the proper wave equation from the numerous candidate equations, we use the physics-informed information criterion (PIC) \cite{xu2022discovery}, which simultaneously considers the evaluation of parsimony and precision, resulting in an equation with physical interpretability. After discovering the structure of the wave equation, a physics-informed neural network (PINN) \cite{raissi2019physics}, which is initialized by the trained network, is trained to identify the corresponding coefficients for every term in the discovered wave equation. We present the discovery of the 2D acoustic equation to demonstrate the potential of D-WE, and its robustness in handling noisy and sparse data.
\section{method}
\subsection{Problem description}
In this work, we consider the general form of a seismic wave equation consisting of:
\begin{equation}\label{eq1}
u_{T} = f\left (\boldsymbol{\Theta}(u);\left[\xi_{i}\right]_{i=1,\cdot\cdot\cdot,n} \right)
\end{equation}
with
\begin{equation}\label{eq2}
\boldsymbol{\Theta}(u)=\left[u, u_x, u_y, u_z, u_{x x}, u_{y y}, u_{z z}, \cdots\right],
\end{equation}
where ${\textit u}_T$ denotes different orders of derivatives of displacement ${\textit u}$ with respect to time ${\textit t}$, e.g., first ($u_t$) or second ($u_{tt}$); ${\Theta}(u)$ refers to the candidate library composed of potential function terms, in which the subscripts represent different orders of derivatives in space; $\left[\xi_i \right]_{i=1,\cdot\cdot\cdot,n}$ denotes the vector of coefficients with size $n$ of the candidates in the library; and $f\left(\cdot \right)$ is a function parameterizing a wave equation with possible contributing terms.

For a specific displacement wavefield data, denoted as $u\left (x_i,y_j,z_k,t_l \right)$, $i=1,\cdot\cdot\cdot, N_x$, $j=1,\cdot\cdot\cdot, N_y$, $k=1,\cdot\cdot\cdot, N_z$, and $l=1,\cdot\cdot\cdot, N_t$, equation (\ref{eq1}) can be expressed as
\begin{equation}\label{eq3}
\left[\begin{array}{c}
u_T\left(x_1, y_1, z_1, t_1\right) \\
u_T\left(x_2, y_1, z_1, t_1\right) \\
\ldots \\
u_T\left(x_{N_x}, y_{N_y}, z_{N_z}, t_{N_t}\right)
\end{array}\right]=\left[\begin{array}{ccc}
u\left(x_1, y_1, z_1, t_1\right) & u_x\left(x_1, y_1, z_1, t_1\right) & \cdots \\
u\left(x_2, y_1, z_1, t_1\right) & u_x\left(x_2, y_1, z_1, t_1\right) & \ldots \\
\ldots & \ldots & \ldots \\
u\left(x_{N_x}, y_{N_y}, z_{N_z}, t_{N_t}\right) & u_x\left(x_{N_x}, y_{N_y}, z_{N_z}, t_{N_t}\right) & \ldots
\end{array}\right]\left[\begin{array}{c}
\xi_1 \\
\ldots \\
\xi_n
\end{array}\right].
\end{equation}
We can see that the linear system in (\ref{eq3}) has $N_x \cdot N_y \cdot N_z \cdot N_t$ equations and $n$ unknown coefficients. In most cases, $N_x \cdot N_y \cdot N_z \cdot N_t \gg n$ holds, which implies equation (\ref{eq3}) is an over-determined system. Fortunately, a parsimonious form exists in most seismic wave equations, thus, just a few function terms have nonzero coefficients. 

 The objective of a data-driven discovery of a wave equation is to identify the closed form of $f\left(\cdot \right)$, that is, find the linear combination of function terms in ${\Theta}(u)$ and corresponding vector of coefficients $\left[\xi_i \right]_{i=1,\cdot\cdot\cdot,n}$. Actually, these two problems are mutually dependent. On the one hand, once we identify which coefficients are nonzero from the coefficients $\left[\xi_i \right]_{i=1,\cdot\cdot\cdot,n}$, the structure of the correct equation is also determined. On the other hand, if we discover the structure of the equation, then how to obtain the nonzero coefficients becomes a simple problem of solving a linear equation.

\subsection{The Neural network}
As mentioned, the problem of discovering the seismic wave equation is given by equation (\ref{eq3}). However, how can we obtain the derivative terms on the LHS and RHS when only the displacement wavefield is known? A feasible option is to use finite difference and other numerical methods to calculate time and spatial derivatives. Although numerical methods are efficient, the observations need to be on a regular grid and have high signal-to-noise ratio for accurate numerical derivatives, which is usually difficult to guarantee, especially for field data.

\begin{figure*}
\vspace{2mm}
\centering
\includegraphics[width=0.6\textwidth]{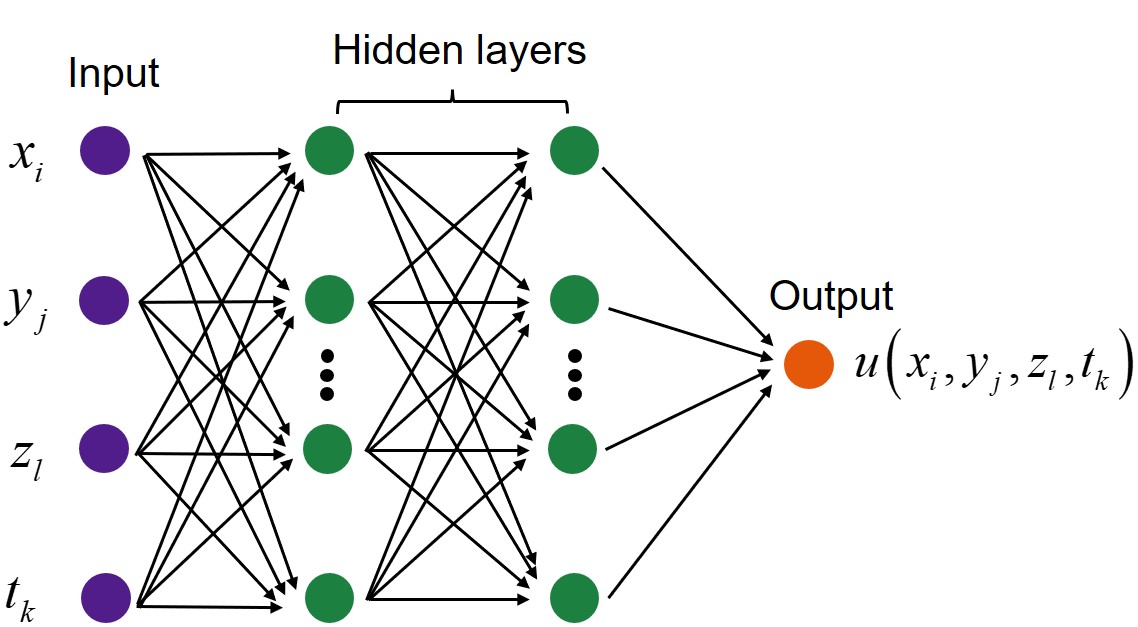}
\caption{Structure of deep fully connected backpropagation NN.}
\label{fig1}
\end{figure*}

In contrast, NNs \change{has proven its}{have proven their} robustness in representing noisy data \cite{xu2022discovery}. Hence, an alternative solution is that we can skillfully use automatic differentiation of the NN to calculate the derivatives during the process of backpropagation. For this purpose, we only need to train a deep fully connected backpropagation NN (shown in Figure \ref{fig1}) using the following loss function
\begin{equation}\label{eq4}
\mathcal{L}(\theta) =\frac{1}{N_x N_y N_z N_t} \sum_{i=1}^{N_x} \sum_{j=1}^{N_y} \sum_{k=1}^{N_z} \sum_{l=1}^{N_t} \left [u(x_i,y_j,z_k,t_l)-\mathrm{NN}(x_i,y_j,z_k,t_l;\theta) \right]^2,
\end{equation}
where $\theta$ denotes the NN trainable parameters, and the corresponding inputs to the network are the spatial-temporal locations $\left (x_i,y_j,z_k,t_l \right)$, to approximate the displacement wavefield $u\left (x_i,y_j,z_k,t_l \right)$ and its derivatives. In real applications, the wavefield (i.e., labels) \remove{can} are given by the recorded data. However, for this analysis of the approach, we will simulate the data. 

Since the inputs to the network are mutually independent position coordinates, the \change{often}{typical} requirement that \change{observation}{observed} data must be collected on a regular grid is not need here. Another advantage of this implementation is that the trained NN can be further used to \change{generate meta-data, or in other words, interpolate data}{predict the wavefield at certain spatial locations where we do not have observations. The predicted wavefields are referred as meta-data, which can further expand the data volume to us, thereby assisting in the discovery of the wave equation}. Furthermore, this trained NN will serve as the initialization for the PINN, significantly reducing the training cost required to accurately determine the equation structure using PIC and identify the coefficients\remove{ using PINN}. We will elaborate on this in detail later on.

\subsection{Genetic algorithm}
\begin{figure*}
\vspace{2mm}
\centering
\includegraphics[width=0.8\textwidth]{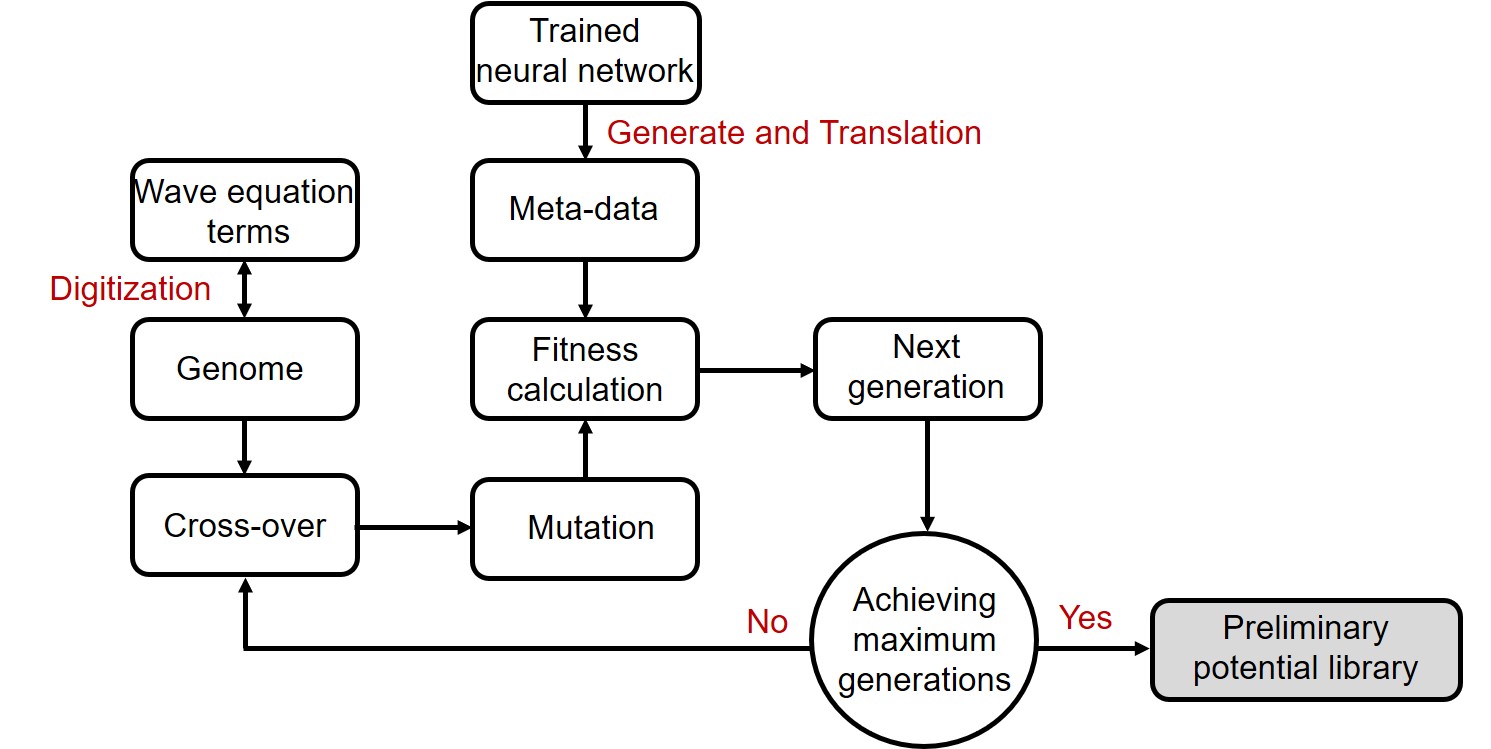}
\caption{The workflow of the genetic algorithm. }
\label{fig2}
\end{figure*}

GA is a type of optimization algorithm inspired by the process of natural selection. Since directly solving equation (\ref{eq3}) is a nondeterministic polynomial time (NP) hard problem with unlimited combinations \cite{xu2022discovery}, we utilize GA to search for an optimal preliminary candidate functional terms from unlimited combinations, which converts the problem to a finite-dimensional problem. The GA does this by creating a population of candidate solutions, each of which represents a different set of functional terms. The fitness of each candidate term is evaluated based on its ability to accurately model the wave behavior of the proposed equation. In our method, GA uses a series of operations, including translation, crossover, mutation, and selection to evolve the population of candidate solutions over multiple generations. The workflow of GA is presented in Figure \ref{fig2}. In the following, we illustrate the steps in detail.

Firstly, we introduce a principle of translation to digitize the structure of potential seismic wave equation to the corresponding genome\remove{, considering the symmetry of the wave equation}. \add{Specifically, we first use numbers, which is defined as the gene, to represent different order of derivatives.} For example,
\begin{equation}\label{eq5}
\text{Gene:} \left\{\begin{array}{l}
{0 \Leftrightarrow u} \\
{1 \Leftrightarrow u_x \quad \text{or} \quad u_y \quad \text{or} \quad u_z \quad \text{or} \quad u_t} \\
{2 \Leftrightarrow u_{xx} \quad \text{or} \quad u_{yy} \quad \text{or} \quad u_{zz} \quad \text{or} \quad u_{tt}} \\
{3 \Leftrightarrow u_{xxx} \quad \text{or} \quad u_{yyy} \quad \text{or} \quad u_{zzz}} \\
\end{array}\right..
\end{equation}
\add{Here, number 0 represents displacement/pressure wavefield $u$; number 1, 2, and 3 are used to encode the first, second, and third order spatial derivatives of the displacement/pressure wavefield, respectively. Also, we use numbers 1 and 2 to represent first and second order time derivatives of the displacement/pressure wavefield, respectively. Then, we combine some genes to form gene modules, which can be utilized to define function terms. For example, the function terms in the LHS are represented as}
\begin{equation}\label{eq6}
\text{Gene module:} \left\{\begin{array}{l}
{[1] \Leftrightarrow u_t} \\
{[2] \Leftrightarrow u_{tt}} 
\end{array}\right.,
\end{equation}
\add{while the function terms in the RHS have the form}
\begin{equation}\label{eq7}
\text{Gene module:} \left\{\begin{array}{l}
{[2] \Leftrightarrow u_{x x}+u_{y y}+u_{z z}} \\
{[0,3] \Leftrightarrow uu_{x x x}+uu_{y y y}+uu_{z z z}}
\end{array}\right.,
\end{equation}
\add{It's noted that for most wave equations, the LHS of the equation is expressed as a first- or second-order time derivative. Hence, we only consider these two cases here. Meanwhile, we have to emphasize that we consider the symmetry of the wave equation in the RHS of the equation. For example, the gene module $[2]$, which is different from gene 2, not only represents the second-order spatial derivatives of the displacement/pressure wavefield with respect to spatial variables $x$, but also signifies the second-order spatial derivatives with respect to the spatial variables $y$ and $z$. The spatial derivatives of the displacement/pressure wavefield with respect to these three spatial variables are combined through addition. Additionally, when a gene module has multiple genes, it represents the multiplication of the corresponding genes, that is, $[0, 3]$ denotes $uu_{x x x}+uu_{y y y}+uu_{z z z}$. The combination of gene modules is regarded as the genome, which includes the LHS and RHS terms in the potential wave equation. We assume the LHS of wave equation only includes the derivatives with respect to $t$, and the LRS of wave equation consists of the derivatives with respect to spatial variables $x$, $y$, and $z$, which applies to many PDEs. Hence, in the case of the LHS of the equation given by the first-order time derivative, we can use the following digitization to translate the corresponding wave equation:}
\begin{equation}\label{eq8}
\text{Genome: } [1]\{[2],[0,3]\} \Leftrightarrow u_t = u_{x x}+u_{y y}+u_{z z} + uu_{x x x}+uu_{y y y}+uu_{z z z}.
\end{equation}
\add{When the LHS of the equation is given by the second-order time derivative, we can represent the wave equation as follows:}
\begin{equation}\label{eq9}
\text{Genome: } [2]\{[2],[0,3]\} \Leftrightarrow u_{tt} = u_{x x}+u_{y y}+u_{z z} + uu_{x x x}+uu_{y y y}+uu_{z z z}.
\end{equation}
\add{These genomes are presented shown as examples. Similar digitization and encoding can be obtained analogously. We note that the gene modules here are connected by addition. Moreover, we have not digitized the coefficients, such as velocity and density, which are commonly present in the wave equation. Our current focus is discovering the wave equation without prior knowledge of the specific values of these coefficients. The values of the coefficients will be determined using PINN after discovering the equation's structure, as will be introduced in Section 2.5. In equations 8 and 9, we have set the coefficients of the function terms $u_{x x}+u_{y y}+u_{z z}$ and $uu_{x x x}+uu_{y y y}+uu_{z z z}$ to 1 for the sake of illustration. However, the coefficients can take on arbitrary values, and this does not impact our ability to discover the equation's structure. }

\remove{respectively. Here, we use numbers, which is defined as the gene, to represent different order of derivatives. For example, as shown in equations 5 and 6, in the LHS, we use number 1 to encode the first order of time derivatives of the displacement wavefield $u_t$; In the RHS, the zero, first, second, and third orders of spatial derivatives of the displacement wavefield are coded with numbers 0, 1, 2, and 3, respectively. Each term is transformed as a gene module, which is composed of genes. For example, the gene module $[1,2]$ refers to $u_x u_{x x}+u_y u_{y y}+u_z u_{z z}$. The combination of gene modules is regarded as the genome, denoted here as ${[1]\{[1,2],[0,0,3]\}}$, which includes the LHS and RHS terms in the potential wave equation. Similar digitization and encoding can be obtained by analogy}

\begin{figure*}
\vspace{2mm}
\centering
\includegraphics[width=1\textwidth]{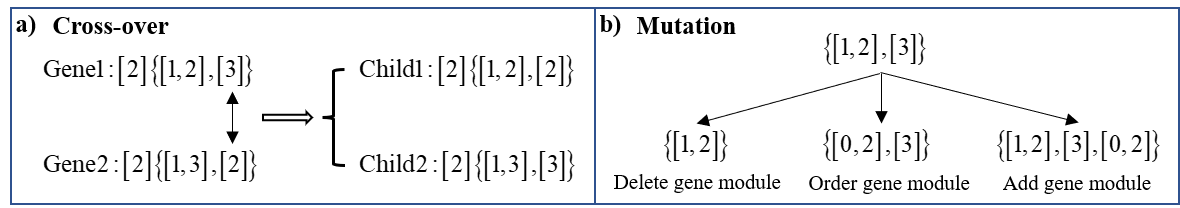}
\caption{An illustration of the process of cross-over and mutation. }
\label{fig3}
\end{figure*}

Subsequently, cross-over and mutation are conducted under a certain probability to obtain next generation candidates. Cross-over means swapping parts of gene modules of two genomes to generate their children (see Figure \ref{fig3}a). Following the cross-over, mutation produces new genes, containing add, delete, and order genes (see Figure \ref{fig3}b). It should be emphasized that crossover and mutation are only applied to the RHS of the equation, whereas the LHS searches for the time derivative order. This is reasonable for most wave equations. 

After mutation, we need to measure the quality of the genome and then perform the selection process. The measurement index is computed by a fitness function as follows:
\begin{equation}\label{eq10}
\mathcal{F} = \frac{1}{N}\sum\left(e q u^L-e q u_i^R \xi_i\right)^2 +\epsilon \cdot len(genome),
\end{equation}
where $N$ denotes all observation samples, $equ^L$ denotes the LHS function terms of the candidate wave equation, $equ_i^R$ represent $i$ th function term in the RHS, and the corresponding coefficients $\xi_i$ are calculated using singular value decomposition (SVD). \add{It is worth emphasizing that in this section and the next section, we are utilizing the SVD method to calculate the coefficients corresponding to each function term. Although it may not achieve absolute precision, it can be relied upon.} To avoid redundancy in the discovery equation, we use an $l_0$ penalty on the number of terms in the discovered equation. Here, $len(genome)$ denotes the length of the genome, and $\epsilon$ is a hyperparameter. In general, as $\epsilon$ increases, the equation becomes more concise. Conversely, as $\epsilon$ decreases, the equation exhibits a more complex structure. 

Once we obtain the fitness of all potential candidate equations, we can select the genomes that better describes the wave propagation system. In our case, the best half of the children are selected as the next generation of parents, and all others genomes are replaced by new random genomes. The process of cross-over, mutation, and selection is repeated for the new generation. When a certain predefined iteration is reached, the preliminary library with a few terms in the last generation is reserved. For this preliminary library, the combinations of all candidate function terms are countable, which is useful to evaluate each combination to further determine the equation. To do this, in the next section, we will use the PIC algorithm \cite{xu2022discovery} to discover the accurate structure of the wave equation from the preliminary library.  

\subsection{Physics-informed information criterion}

\begin{figure*}
\vspace{2mm}
\centering
\includegraphics[width=0.6\textwidth]{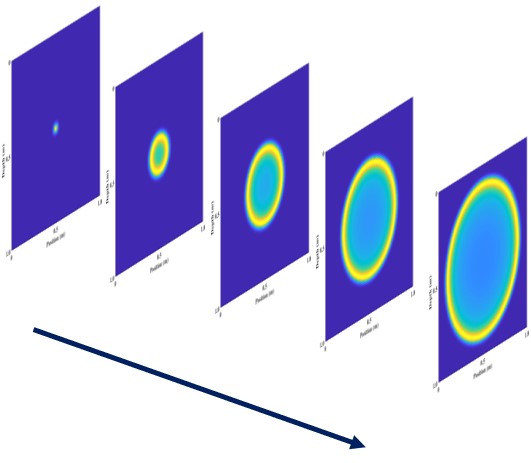}
\caption{An illustration of moving horizon.}
\label{fig4}
\end{figure*}

The PIC algorithm involves two types of measurements: redundancy and physical losses. Redundancy loss is used to measure the parsimony of the proposed equation and is based on the idea that the coefficients of redundant terms are unstable \change{in the moving horizon}{when applied to the observed data on moving windows of a given time step} \cite{lejarza2022data}. Therefore, we can \add{utilize this technique, namely the moving horizon, to} calculate the average variation in coefficients for each combination to obtain the redundancy loss\remove{, using the moving horizon technique}. As shown in the Figure \ref{fig4}, the smooth wavefield snapshots generated by the NN at different times are divided into $N_h$ overlapping horizons $T_i$ ($i = 1, \cdot \cdot \cdot, N_h$). The $T_i$ is denoted as the wavefield snapshots within a time range, \change{like}{such as} $[t_{min}^{'} + i\Delta t, \frac{1}{2}(t_{min}^{'} + t_{max}^{'})+i\Delta t]$, where $t_{min}^{'}$ and $t_{max}^{'}$ represent the the minimum and maximum of the time domain of the generated snapshots, respectively, and $\Delta t$ denotes the length of horizons. For a candidate combination ${\Theta}_j$ (i.e., potential wave equation), the corresponding vector of coefficients $\xi_j^i$ in horizon $T_i$ can be obtained solving equation ${equ}_{i,j}^L - \xi_j^i \cdot {equ}_{i,j}^R = 0$, where ${equ}_{i,j}^L$ and ${equ}_{i,j}^R$ are the values of the LHS and RHS terms for a potential wave equation ${\Theta}_j$ in horizon $T_i$, respectively. For the combination ${\Theta}_j$, when we obtain all the coefficient vectors in overlapping horizons $T_i$, $i = 1, \cdot \cdot \cdot, N_h$, we can calculate the corresponding redundancy loss as follows:
\begin{equation}\label{eq11}
\mathcal{L}_r(\Theta_j) = \frac{1}{N_\mathrm{term}}\sum_{k=1}^{N_\mathrm{term}}\frac{\sigma_{j,k}}{\mu_{j,k}},
\end{equation}
where $N_\mathrm{term}$ denotes the number of terms, $\sigma_{j,k}$ and $\mu_{j,k}$ represent the standard deviation and mean, respectively, of the $N_h$ different coefficients over the overlapping horizons corresponding to the $k~th$ function term in the candidate combination ${\Theta}_j$. As can be appreciated in equation (\ref{eq11}), the accurate terms are stable in the moving horizons, that is, the coefficients have small standard deviations, resulting in a small redundancy loss. In contrast, the coefficients of redundant terms exhibit a large degree of variation in different horizons, due to the need to compensate for \change{uneven errors caused by different noise in different horizons}{errors caused by noise, which could be different in different horizons}.

The physical loss, which is based on PINN \cite{raissi2019physics}, is presented to evaluate the accuracy of the discovered wave equation. Here, we need first train a PINN, which maintains the same architecture as NN shown in Figure \ref{fig1} and is initialized by the NN parameters $\theta$, while the loss function has the form of
\begin{equation}\label{eq12}
\mathcal{L}_\mathrm{PINN}(\theta) = \lambda_{d} \mathrm{MSE}_{d} + \lambda_{p}\mathrm{MSE}_{p}
\end{equation}
with the data loss
\begin{equation}\label{eq13}
\mathrm{MSE}_{d} = \frac{1}{N_x N_y N_z N_t} \sum_{i=1}^{N_x} \sum_{j=1}^{N_y} \sum_{k=1}^{N_z} \sum_{l=1}^{N_t} \left [u(x_i,y_j,z_k,t_l)-\mathrm{PINN}(x_i,y_j,z_k,t_l;\theta) \right]^2
\end{equation}
and the PDE loss
\begin{equation}\label{eq14}
\mathrm{MSE}_{p} = \frac{1}{N_x^{'} N_y^{'} N_z^{'} N_t^{'}} \sum_{i=1}^{N_x^{'}} \sum_{j=1}^{N_y^{'}} \sum_{k=1}^{N_z^{'}} \sum_{l=1}^{N_t^{'}} \left [equ_L^{'}(x_i^{'},y_j^{'},z_k^{'},t_l^{'})-\xi^{'} \cdot equ_R^{'}(x_i^{'},y_j^{'},z_k^{'},t_l^{'};\theta) \right]^2,
\end{equation}
where $\lambda_d$ and $\lambda_p$ are hyperparameters, which control the contribution of data and PDE losses to the total loss, respectively. Here, the data loss comes from the average squared error (MSE) between the observed data and the predicted one from PINN, whereas the PDE loss is obtained by measuring the MSE between the LHS $equ_L^{'}$ and RHS $\xi^{'} \cdot equ_R^{'}$ terms of the potential wave equation, which is calculated on the meta-data $(x_i^{'},y_j^{'},z_k^{'},t_l^{'})$ generated from the NN. It should be emphasized that the coefficients $\xi^{'}$ are deduced by computing $equ_L^{'} - \xi^{'} \cdot equ_R^{'} = 0$ during each training process. 

After training \add{the} PINN, the physical loss for the potential wave equation (i.e., candidate combination ${\Theta}_j$) can be calculated as follows
\begin{equation}\label{eq15}
\mathcal{L}_{p}(\Theta_j) = \left ( \frac{1}{N_x^{'} N_y^{'} N_z^{'} N_t^{'}} \sum_{i=1}^{N_x^{'}} \sum_{j=1}^{N_y^{'}} \sum_{k=1}^{N_z^{'}} \sum_{l=1}^{N_t^{'}} \left [\hat u^\mathrm{PINN}(x_i^{'},y_j^{'},z_k^{'},t_l^{'}) - \hat u^\mathrm{NN}(x_i^{'},y_j^{'},z_k^{'},t_l^{'}) \right]^2 \right )^{\frac{1}{2}},
\end{equation}
where $\hat u^\mathrm{PINN}$ and $\hat u^\mathrm{NN}$ refer to the normalized output of the meta-data predicted by PINN and NN, respectively, which is determined by:
\begin{equation}\label{eq16}
\setlength{\abovedisplayskip}{3pt}
\setlength{\belowdisplayskip}{3pt}
\hat u^\mathrm{PINN} = \frac{u^\mathrm{PINN} - u_\mathrm{min}}{u_\mathrm{max} - u_\mathrm{min}},
\end{equation}
\begin{equation}\label{eq17}
\setlength{\abovedisplayskip}{3pt}
\setlength{\belowdisplayskip}{3pt}
\hat u^\mathrm{NN} = \frac{u^\mathrm{NN} - u_\mathrm{min}}{u_\mathrm{max} - u_\mathrm{min}},
\end{equation}
where $u_\mathrm{max}$ and $u_\mathrm{min}$ denote the maximum and minimum of the observation data, respectively. The utilization of such a form of physical loss is based on the following fact: when physical constraints are consistent with the data, the predicted results will exhibit significant improvements (see Equation (\ref{eq12})); However, if the physical constraints and data are not parallel, the performance of PINN will decrease. As a result, if the underlying wave equation can effectively describe the wavefield data, the predicted results of the trained PINN is closer to the NN's output, which is relatively accurate. Hence, the physical loss will be very small.

 For each candidate combination ${\Theta}_j$, the PIC is obtained by multiplying the calculated redundancy and physical losses as follows:
 \begin{equation}\label{eq18}
\mathrm{PIC}(\Theta_j) = \mathcal{L}_{r}(\Theta_j) \cdot \mathcal{L}_{p}(\Theta_j).
\end{equation}
It is worth noting that the PIC is not performed for all possible combinations in the preliminary potential library, as calculating physical loss is time-consuming. Since the computational cost for redundancy loss is cheap, we first derive all redundancy loss and select top $N_b$ combinations with smaller redundancy loss. Following that, we perform the PINN training on the $N_b$ combinations and further combine redundancy and physical losses to present PIC. Afterwards, we will discover the correct structure of wave equation with the smallest PIC. 

\subsection{Identifying coefficients}
Although we assume that the general structure of the wave equation has been obtained, we still need to determine the coefficients. Certainly, we can obtain the coefficients by directly solving \change{an often}{a typically} overdetermined system of linear equations (\ref{eq3}). For example, we use the SVD method to calculate the coefficients in equations (\ref{eq10}) and (\ref{eq14}). However, this solution is not exact, especially for noisy data. In contrast, PINN provides a reliable framework to identify the coefficients. Hence, we use the PINN to obtain the values of the coefficients, which is also initialized by the previously trained network NN (Figure \ref{fig1}), while the loss function is reset to
\begin{equation}\label{eq19}
\mathcal{L}(\theta, \xi)= \frac{1}{N_x^{'} N_y^{'} N_z^{'} N_t^{'}} \sum_{i=1}^{N_x^{'}} \sum_{j=1}^{N_y^{'}} \sum_{k=1}^{N_z^{'}} \sum_{l=1}^{N_t^{'}} \left [equ_L^{'}(x_i^{'},y_j^{'},z_k^{'},t_l^{'})-\xi \cdot equ_R^{'}(x_i^{'},y_j^{'},z_k^{'},t_l^{'};\theta,\xi) \right]^2.
\end{equation}
Here, the coefficients $\xi$ are not the output of PINN, and thus, we define it as additional trainable parameters of PINN, which is updated along with network parameters $\theta$ to minimize the loss function. We initialize the values for the trainable coefficients $\xi$ from solving the system of linear equations (\ref{eq3}). Although initially it may not be accurate, it can help PINN converge faster than starting with random initializations. Once we identify the exact coefficients, we can combine it with the corresponding function terms to obtain the general form of the discovered wave equation.
\section{Numerical~Examples}
To verify the feasibility and effectiveness of D-WE, we present an example in discovering the 2D acoustic wave equation:
\begin{equation}
\setlength{\abovedisplayskip}{3pt}
\setlength{\belowdisplayskip}{3pt}
u_{t t}=v^2\left(u_{x x}+u_{z z}\right),
\end{equation}
where we assume that the body force is absent, and $v$ denotes velocity. We consider wave propagation in a homogeneous medium and utilizes \remove{the}finite difference\add{s} (FD) to generate the dataset. The medium, which we assume has a velocity 2 km/s, is discretized along 101 grid points in both $x$ and $z$ directions with a grid spacing of 10 m. We collect 121 snapshots of the pressure wavefield with a time interval of 2 ms from zero to 0.24 seconds. The wavefield is initiated by an isotropic Gaussian function at the center of the model given by
\begin{equation}
\setlength{\abovedisplayskip}{3pt}
\setlength{\belowdisplayskip}{3pt}
u(i, j, 0)=\exp \left(-0.2 *\left[(i-51)^2+(j-51)^2\right]\right), \quad i, j=1, \cdots, 101,
\end{equation}
at time zero. In this test, the NN has three hidden layers with 50 neurons in each layer. \add{Refer to} \cite{xu2022discovery}, the activation function is \add{set as} a sine function. The maximum population size of genomes is 400, the maximum number of generations is taken as 100. \add{The} NN and PINN are trained by using an Adam optimizer \cite{kingma2014adam}. The hyperparameters $\epsilon$ (equation (\ref{eq10})), $\lambda_d$, and $\lambda_p$ (equation (\ref{eq12})) are set to $10^{-6}$, $1$, and $0.01$, respectively.

\add{We first provide an example to illustrate the process of our method in discovering the acoustic wave equation from observed pressure wavefields. We randomly select ${20\%}$ subsets from the complete volume of observed pressure wavefields to train the NN and then utilize them to discover the equation. The NN is trained for $30000$ iterations. We simultaneously consider cases where the LHS of the equation is given by a first-order and a second-order time derivatives. We generate the initial library on the LRS of the equation as $\{[0, 1, 3], [1, 1]\}$, corresponding to the form $uu_xu_{xxx} + uu_zu_{zzz} + u_x^2 + u_z^2$. By utilizing the GA (as illustrated in Figure 2), the initial library will evolve to produce a overcomplete library, including a lot of candidate function terms. In our case, we limit the number of candidate function terms to 400, which constitutes the maximum population size of genomes. We list the optimal genomes at some generations, where Tables 1 and 2 correspond to the equations with first-order and second-order of time derivatives on the LHS, respectively. The first column represents the number of generation of evolution, the second column indicates the optimal genome and the corresponding translated form of the potential equation, and the final column represents their corresponding fitness scores.

From Tables 1 and 2, we can see that as with the progress in evolution, the optimal genome tends to stabilize. For example, when the LHS of the equation is a first-order of time derivative, the optimal genome is $\{[2], [0, 0, 2], [0, 1, 2], [0, 1, 3]\}$, while when the the LHS of the equation corresponds to a second-order of derivative, the optimal genome is $\{[2], [0, 0, 2], [0, 1, 2], [0, 1, 3]\}$. However, if we are to stop here and choose the equation form based on fitness scores, it would be $\{[2], [0, 0, 2], [0, 1, 2], [0, 1, 3]\}$. Certainly, this does not match the accurate form of the acoustic wave equation. Therefore, as stated earlier, we consider the optimal genome from the GA at the maximum number of generations as a preliminary library. The combinations from this preliminary library can be countable. We can select arbitrary terms to form a potential structure of the wave equation and then utilize a more accurate PIC metrics to determine the exact structure of the equation from all combinations. Tables 3 and 4 display the potential wave equations with the lowest 5 PIC metrics, corresponding to the first-order and second-order time derivatives on the LHS, respectively. By comparing Tables 3 and 4, we can see that the potential wave equation in the form of $u_{tt} = \xi_1(u_{xx} + u_{zz})$ has the lowest PIC values, and as a result, is ultimately identified as the discovered equation structure.

Compared with the form of the acoustic wave equation, it is demonstrated that the equation structure, which is directly discovered from the observed pressure wavefield, is consistent with the corresponding acoustic wave equation. Once we establish the accurate structure of the equation, we further utilize PINN to optimize the coefficient $\xi_1$ for the terms $(u_{xx} + u_{zz})$. Ultimately, we obtain the coefficient for the equation as $3.989 ~ \mathrm{km^2/s^2}$, which has a negligible error with the true value $v^2 = 4 ~ \mathrm{km^2/s^2}$. Thus, the discovered general form of the equation is $u_{tt} = 3.989(u_{xx} + u_{zz})$. It is worth emphasizing that apart from the training time of the NN, the entire discovery process is highly efficient, costing only a total of 198 seconds.}

\begin{table}
\centering
\caption{The evolution process of the preliminary library when the left-hand side of the equation is the first-order time derivative.}
\renewcommand\arraystretch{1.5}
\setlength{\tabcolsep}{25pt}
\begin{tabular}{ccc}
    \hline
     \multirow{3}{*} \text{Number of}  & \text{Genome and } & \text {Fitness}  \\
     \text{generations}  & \text{Translation} & \\
    \hline
     $1$   & $\text {Genome: } [1]{\{[1, 1], [0, 1, 3]}\}$  &  $135.62$ \\
           & $\text {Translation: } \it{u}_{t}= \xi_1(u_{x}^2+u_{z}^2)+\xi_2(uu_xu_{xxx}+uu_zu_{zzz})$ &  \\
    \hline
    $20$     & $\text {Genome: } [1]{\{[1, 1, 2], [0, 2, 2], [0, 1, 1], [1, 1]}\}$  &  $110.37$ \\
         & $\text {Translation: } \it{u}_{t}= \xi_1(u_{x}^2u_{xx}+u_{z}^2u_{zz})+\xi_2(uu_{xx}^2+uu_{zz}^2)$ &  \\
          & $ \quad \quad \quad \quad \quad +\xi_3(uu_{x}^2+uu_{z}^2)+\xi_4(u_{x}^2+u_{z}^2)$ &  \\
    \hline
    $40$     & $\text {Genome: } [1]{\{[0, 0], [0, 2], [0, 1, 1], [1, 1]}\}$  &  $47.27$ \\
         & $\text {Translation: } \it{u}_{t}= \xi_1u^2+\xi_2(uu_{xx}+uu_{zz})$ &  \\
          & $ \quad \quad \quad \quad \quad +\xi_3(uu_{x}^2+uu_{z}^2)+\xi_4(u_{x}^2+u_{z}^2)$ &  \\
    \hline
    $60$     & $\text {Genome: } [1]{\{[0, 0], [0, 2], [0, 1, 1], [1, 1]}\}$  &  $47.27$ \\
         & $\text {Translation: } \it{u}_{t}= \xi_1u^2+\xi_2(uu_{xx}+uu_{zz})$ &  \\
          & $ \quad \quad \quad \quad \quad +\xi_3(uu_{x}^2+uu_{z}^2)+\xi_4(u_{x}^2+u_{z}^2)$ &  \\
    \hline
    $80$     & $\text {Genome: } [1]{\{[0, 0], [0, 2], [0, 1, 1], [1, 1]}\}$  &  $47.27$ \\
         & $\text {Translation: } \it{u}_{t}= \xi_1u^2+\xi_2(uu_{xx}+uu_{zz})$ &  \\
          & $ \quad \quad \quad \quad \quad +\xi_3(uu_{x}^2+uu_{z}^2)+\xi_4(u_{x}^2+u_{z}^2)$ &  \\
    \hline
    $100$     & $\text {Genome: } [1]{\{[0, 0], [0, 2], [0, 1, 1], [1, 1]}\}$  &  $47.27$ \\
         & $\text {Translation: } \it{u}_{t}= \xi_1u^2+\xi_2(uu_{xx}+uu_{zz})$ &  \\
          & $ \quad \quad \quad \quad \quad +\xi_3(uu_{x}^2+uu_{z}^2)+\xi_4(u_{x}^2+u_{z}^2)$ &  \\
    \hline
\end{tabular}
\label{tab1}
\end{table}

\begin{table}
\centering
\caption{The evolution process of the preliminary library when the left-hand side of the equation is the second-order time derivative.}
\renewcommand\arraystretch{1.5}
\setlength{\tabcolsep}{25pt}
\begin{tabular}{ccc}
    \hline
     \multirow{3}{*} \text{Number of }  & \text{Genome and } & \text {Fitness}  \\
     \text{generations}  & \text{Translation} & \\
    \hline
     $1$   & $\text {Genome: } [2]{\{[2], [0, 0, 2], [0, 1, 1]}\}$  &  $47880.75$ \\
           & $\text {Translation: } \it{u}_{tt}= \xi_1(u_{xx}+u_{zz})+\xi_2(u^2u_{xx}+u^2u_{zz})$ &  \\
           & $+ \xi_3(uu_{x}^2+uu_{z}^2)$ &  \\
    \hline
    $20$     & $\text {Genome: } [2]\{[2], [0, 0, 2], [0, 1, 3]\}$  &  $47748.91$ \\
           & $\text {Translation: } \it{u}_{tt}= \xi_1(u_{xx}+u_{zz})+\xi_2(u^2u_{xx}+u^2u_{zz})$ &  \\
           & $+ \xi_3(uu_{x}u_{xxx}+uu_{z}u_{zzz})$ &  \\
    \hline
    $40$     & $\text {Genome: } [2]\{[2], [0, 0, 2], [0, 1, 2], [0, 1, 3]\}$  &  $47730.40$ \\
           & $\text {Translation: } \it{u}_{tt}= \xi_1(u_{xx}+u_{zz})+\xi_2(u^2u_{xx}+u^2u_{zz})$ &  \\
           & $+ \xi_3(uu_{x}u_{xx}+uu_{z}u_{zz})+ \xi_4(uu_{x}u_{xxx}+uu_{z}u_{zzz})$ &  \\
    \hline
    $60$     & $\text {Genome: } [2]\{[2], [0, 0, 2], [0, 1, 2], [0, 1, 3]\}$  &  $47730.40$ \\
           & $\text {Translation: } \it{u}_{tt}= \xi_1(u_{xx}+u_{zz})+\xi_2(u^2u_{xx}+u^2u_{zz})$ &  \\
           & $+ \xi_3(uu_{x}u_{xx}+uu_{z}u_{zz})+ \xi_4(uu_{x}u_{xxx}+uu_{z}u_{zzz})$ &  \\
    \hline
    $80$     & $\text {Genome: } [2]\{[2], [0, 0, 2], [0, 1, 2], [0, 1, 3]\}$  &  $47730.40$ \\
           & $\text {Translation: } \it{u}_{tt}= \xi_1(u_{xx}+u_{zz})+\xi_2(u^2u_{xx}+u^2u_{zz})$ &  \\
           & $+ \xi_3(uu_{x}u_{xx}+uu_{z}u_{zz})+ \xi_4(uu_{x}u_{xxx}+uu_{z}u_{zzz})$ &  \\
    \hline
    $100$     & $\text {Genome: } [2]\{[2], [0, 0, 2], [0, 1, 2], [0, 1, 3]\}$  &  $47730.40$ \\
           & $\text {Translation: } \it{u}_{tt}= \xi_1(u_{xx}+u_{zz})+\xi_2(u^2u_{xx}+u^2u_{zz})$ &  \\
           & $+ \xi_3(uu_{x}u_{xx}+uu_{z}u_{zz})+ \xi_4(uu_{x}u_{xxx}+uu_{z}u_{zzz})$ &  \\
    \hline
\end{tabular}
\label{tab2}
\end{table}

\begin{table}
\centering
\caption{The potential wave equations and the corresponding PIC when the left-hand side of the equation is the first-order time derivative.}
\renewcommand\arraystretch{1.5}
\setlength{\tabcolsep}{20pt}
\begin{tabular}{cc} 
    \hline
    \text{Potential wave equation} & \text{PIC } \\
    \hline
    ${u}_{t}=\xi_1u^2+\xi_2(uu_{xx}+uu_{zz})$ & $0.028099$ \\
    \hline
    ${u}_{t}=\xi_1u^2+\xi_2(uu_{xx}+uu_{zz})+\xi_2(u_{x}^2+u_{z}^2)$ & $0.0093211$ \\
    \hline
    ${u}_{t}=\xi_1(uu_{xx}+uu_{zz})$ & $0.0075959$ \\
    \hline
    ${u}_{t}=\xi_1u^2+\xi_4(u_{x}^2+u_{z}^2)$ & $0.011005$ \\
    \hline
    ${u}_{t}=\xi_1u^2+\xi_2(uu_{xx}+uu_{zz})+\xi_3(uu_{x}^2+uu_{z}^2)+\xi_4(u_{x}^2+u_{z}^2)$ & $0.011924$ \\
    \hline
\end{tabular}
\label{tab3}
\end{table}

\begin{table}
\centering
\caption{The potential wave equations and the corresponding PIC when the left-hand side of the equation is the second-order time derivative.}
\renewcommand\arraystretch{1.5}
\setlength{\tabcolsep}{20pt}
\begin{tabular}{cc} 
    \hline
    \text{Potential wave equation} & \text{PIC } \\
    \hline
    ${u}_{tt}=\xi_1(u_{xx}+u_{zz})$ & $0.000187$ \\
    \hline
    ${u}_{tt}=\xi_1(u^2u_{xx}+u^2u_{zz})$ & $0.02694$ \\
    \hline
    ${u}_{tt}=\xi_2(uu_{x}u_{xxx}+uu_{z}u_{zzz})$ & $0.035408$ \\
    \hline
    ${u}_{tt}=\xi_1(u^2u_{xx}+u^2u_{zz})+ \xi_2(uu_{x}u_{xxx}+uu_{z}u_{zzz})$ & $0.040359$ \\
    \hline
    ${u}_{tt}=\xi_1(uu_{x}u_{xx}+uu_{z}u_{zz})+ \xi_2(uu_{x}u_{xxx}+uu_{z}u_{zzz})$ & $0.050408$ \\
    \hline
\end{tabular}
\label{tab4}
\end{table}

\begin{figure*}
\vspace*{2mm}
\centering
\includegraphics[width=1\textwidth]{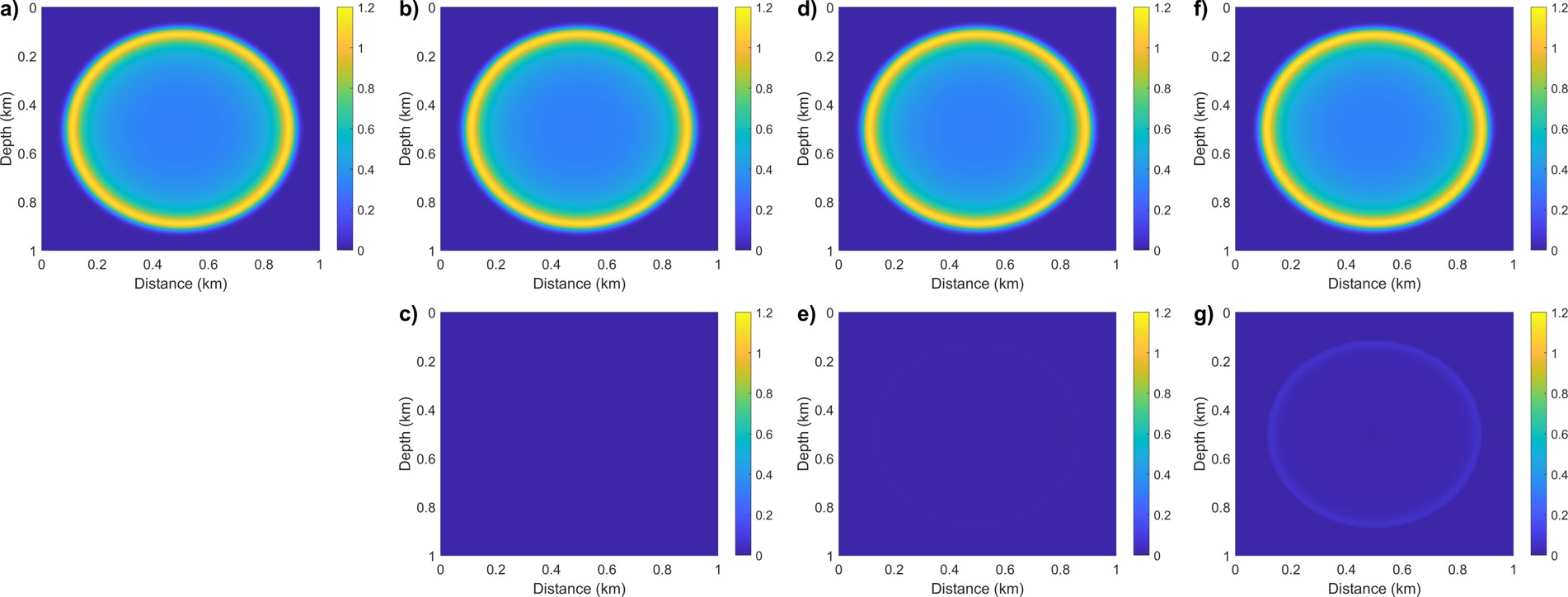}
\caption{Comparison of wavefield snapshots simulated by the accurate acoustic wave equation and the discovered equation with different data volumes. (a) Ground truth comes from accurate acoustic wave equation. (b), (d), and (f) correspond to the discovered equations derived from $60 \%$, $5\%$, and $1\%$ volume data, respectively, and their differences from the ground truth are plotted in (c), (e), and (g), respectively.}
\label{fig5}
\end{figure*}

We then validate the performance of D-WE for discovering the acoustic wave equation \change{with clean data, including the complete (at all grid points), as well as subset volumes pressure wavefield points.}{on sparse observations. We randomly select a subset of the complete grid point measurements of pressure wavefields, including $60\%$, $20\%$, $5\%$, $1\%$, and $0.5\%$ of all grid points, and compare results with the discovered form from the complete pressure wavefield points.} The results are shown in Table \ref{tab5}, where the error represents relative error between coefficients, which is defined as $\left|\xi_{i}-\xi_{i}^{true} \right|/\xi_{i}^{true}$ in percent. As we can see in the table, D-WE enables the discovery of the structure of the acoustic wave equation, even for extremely sparse data (e.g., 0.5\% of all grid points). In most cases, the identified coefficients are almost exactly close to true ones, that is, the square of velocity (in this example equals 4), and only for extremely sparse data, the accuracy of the coefficients is slightly reduced, which is acceptable. We employ FD algorithms to numerically solve discovered equations derived from $60 \%$, $5\%$, and $1\%$ volume data, respectively, and compare the resulting wavefields with those obtained from simulating accurate acoustic wave equation, as depicted in Figure \ref{fig5}. As seen in Figure \ref{fig5}b-e, it is evident that the wavefield snapshots derived from the discovered equation exhibits a remarkably close resemblance to those obtained from accurate acoustic wave equation (Figure \ref{fig5}a) used in generating the observations. Furthermore, even for extremely sparse data, the discovered equation demonstrates a high degree of accuracy in simulating seismic wave propagation (see Figure \ref{fig5}f, g), with only negligible differences.

\begin{figure*}
\vspace*{2mm}
\centering
\includegraphics[width=1\textwidth]{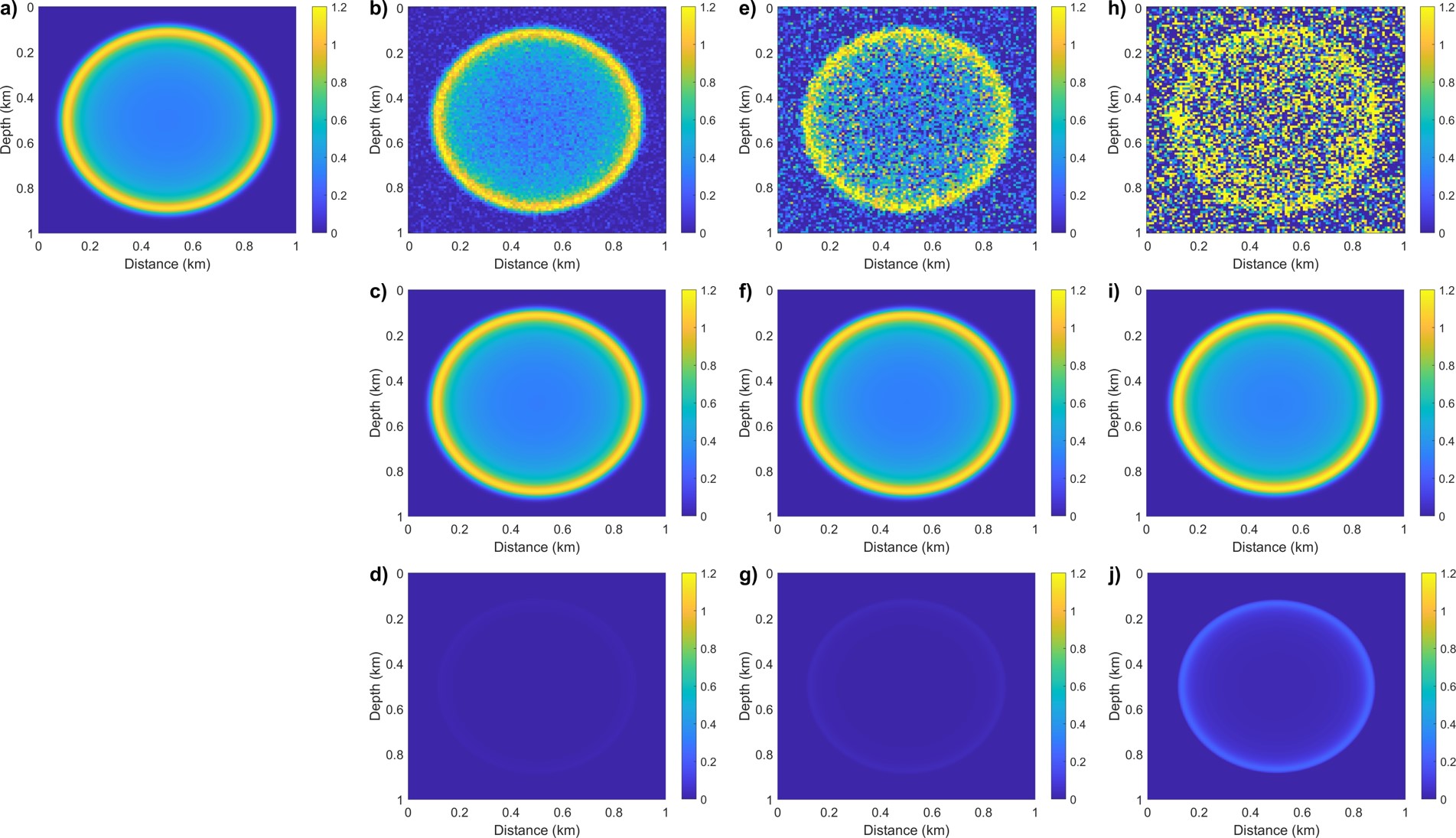}
\caption{Comparison of wavefield snapshots simulated by the accurate acoustic wave equation and the discovered equation with different noise levels. (a) Ground truth comes from accurate acoustic wave equation. (b), (e), and (h) are the noisy wavefield data with noise levels of $25\%$, $100\%$, and $300\%$, respectively, which are obtained by adding noise to the ground truth. (c), (f), and (i) are obtained by solving the discover equations with noise levels of $25\%$, $100\%$, and $300\%$, respectively. (d), (g), and (j) are the corresponding differences with the ground truth.}
\label{fig6}
\end{figure*}

Furthermore, we demonstrate the robustness of D-WE to noisy data, which is presented in Table \ref{tab6}. Here, Gaussian noise is added to the clean data $u$ to obtain the noisy data $\tilde{u}=u+\eta \cdot {std}\left(u \right) \cdot N\left(0,1 \right)$, where $N\left(0,1 \right)$ denotes the standard normal distribution with mean 0 and standard deviation of 1, and $\eta$ is the noise level. The results prove that D-WE is reasonably robust to high levels of noise. Surprisingly, D-WE still accurately discovers the structure of the equation for data with strong noise (e.g., 300\% and 400\% noise level), and limited data. Here, we also numerically solve the discovered equations at noise levels $25 \%$, $100\%$, and $300\%$. Figure \ref{fig6} presents a comparison between the generated wavefield snapshots and their corresponding ground truth (Figure \ref{fig6}a). We can see that our method yields highly accurate equations for observation data with low noise levels (Figure \ref{fig6}b, c, d). As the noise level rises, the wavefield snapshots simulated by the discovered equations show increased signal leakage compared to the ground truth, but the accuracy is still commendable. For observation data with strong noise, such as in Figure \ref{fig6}h, the wavefront is nearly obscured by noise and the continuity is significantly disrupted. However, even in this case, the discovered equation still yields comparable wavefield snapshots to the exact acoustic wave equations, as indicated in Figure \ref{fig6}i, j. 

\begin{figure*}
\vspace*{2mm}
\centering
\includegraphics[width=0.5\textwidth]{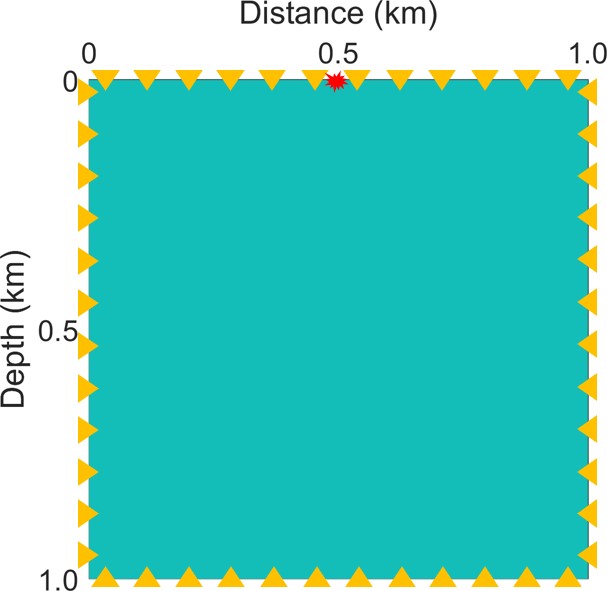}
\caption{Schematic diagram of observation system, where the triangle represents the receivers, and the star represents the position of the center of the isotropic Gaussian function used to initialize the wavefield.}
\label{fig7}
\end{figure*}

To consider more realistic observation systems, we place all receivers on the grid points of the model boundary, also, use the isotropic Gaussian function at the center of the upper surface of the model to initialize the wavefield, which is illustrated in Figure \ref{fig7}. In Table \ref{tab7}, we present the results of the discovered wave equation under this observation system with different noise levels. Remarkably, even under such restricted observation conditions, our method is still able to accurately identify the structure of the equation and maintain robustness to some extent of noise. However, it should be acknowledged that, compared with random observation systems, the restricted boundary observation can lead to the reduced robustness of our method in identifying equation coefficients in the presence of noise. Moreover, when the noise level is high (e.g., noise level = $75\%$), our method may discover an incorrect wave equation. This is a predictable outcome given the highly constrained nature of the observation points.

\begin{table}
\centering
\caption{Test on discovery of a 2D acoustic wave equation with varying subsets of the total observations.}
\renewcommand\arraystretch{1.5}
\setlength{\tabcolsep}{20pt}
\begin{tabular}{ccc}
    \hline
    \text {Volume} {of} {data} & \text { Discovered equation } & \text { Error } \\
    \hline
    $100 \%$ & $\it{u}_{tt}=3.99 \left(u_{xx}+u_{zz}\right)$ & $0.25 \%$ \\
    $60 \%$ & $\it{u}_{tt}=3.999 \left(u_{xx}+u_{zz}\right)$ & $0.025 \%$ \\
    $20 \%$ & $\it{u}_{tt}=3.989 \left(u_{xx}+u_{zz}\right)$ & $0.28 \%$ \\
    $5 \%$ & $\it{u}_{tt}=3.992 \left(u_{xx}+u_{zz}\right)$ & $0.2 \%$ \\
    $1 \%$ & $\it{u}_{tt}=3.941 \left(u_{xx}+u_{zz}\right)$ & $1.48 \%$ \\
    $0.5 \%$ & $\it{u}_{tt}=3.969 \left(u_{xx}+u_{zz}\right)$ & $0.78 \%$ \\
    \hline
\end{tabular}
\label{tab5}
\end{table}

\begin{table}
\centering
\caption{Test on discovery of a 2D acoustic wave equation from data with different noise level.}
\renewcommand\arraystretch{1.5}
\setlength{\tabcolsep}{20pt}
\begin{tabular}{ccc}
    \hline
    \text {Noise} {level} & \text { Discovered equation } & \text { Error } \\
    \hline
    $25 \%$ & $\it{u}_{tt}= 3.985\left(u_{xx}+u_{zz}\right)$ & $0.38 \%$ \\
    $50 \%$ & $\it{u}_{tt}= 3.973\left(u_{xx}+u_{zz}\right)$ & $0.68 \%$ \\
    $100 \%$ & $\it{u}_{tt}=3.97\left(u_{xx}+u_{zz}\right)$ & $0.75 \%$ \\
    $200 \%$ & $\it{u}_{tt}=3.904 \left(u_{xx}+u_{zz}\right)$ & $2.4 \%$ \\
    $300 \%$ & $\it{u}_{tt}=3.76 \left(u_{xx}+u_{zz}\right)$ & $6 \%$ \\
    $400 \%$ & $\it{u}_{tt}=3.517 \left(u_{xx}+u_{zz}\right)$ & $12.08 \%$ \\
    \hline
\end{tabular}
\label{tab6}
\end{table}

\begin{table}
\centering
\caption{Test on discovery of 2D acoustic wave equation from limited observations.}
\renewcommand\arraystretch{1.5}
\setlength{\tabcolsep}{20pt}
\begin{tabular}{ccc}
    \hline
    \text {Noise} {level} & \text { Discovered equation } & \text { Error } \\
    \hline
    $0 \%$ & $\it{u}_{tt}= 3.921\left(u_{xx}+u_{zz}\right)$ & $1.98 \%$ \\
    $25 \%$ & $\it{u}_{tt}= 3.878\left(u_{xx}+u_{zz}\right)$ & $3.05 \%$ \\
    $50 \%$ & $\it{u}_{tt}= 3.706\left(u_{xx}+u_{zz}\right)$ & $7.35 \%$ \\
    $75 \%$ & $\it{u}_{t}=-1.899*10^{-5}\left(uu_{x}u_{xxx}+uu_{z}u_{zzz}\right)$ &  \\
    \hline
\end{tabular}
\label{tab7}
\end{table}

\section{conclusions}
We explored a new methodology to discover wave equations in a data-driven way. This novel implementation, dubbed D-WE, is proposed to directly discover a wave equation from spatial-temporal seismic wavefield observations. D-WE consists of two major components: the neural network (NN) and the genetic algorithm (GA). The NN accepts spatial-temporal locations as inputs of the network to approximate the observation displacement or pressure wavefields, which is used for calculating time and spatial derivatives and producing meta-data. On the other hand, GA serves to generate an expandable candidate function terms library to address the problem of an incomplete initial library. The best wave equation is determined from the candidate library by utilizing a physics-informed information criterion. The corresponding coefficients of each term in the optimal form is identified by PINN, which is initialized by the NN. Test on the discovery of the 2D acoustic wave equation demonstrates that D-WE can identify the correct wave equation and is robust to sparse and noisy wavefield data. This conclusion will hopefully pave the way to us utilizing this approach for the discovery of more exotic wave equations that describe wave propagation more inline with our observations.

\section*{Acknowledgments}
This publication is based on work supported by the King Abdullah University of Science and Technology (KAUST). The authors thank the DeepWave sponsors for supporting this research. We thank Hao Xu for his valuable comments and suggestions. \\

\section*{Data availability statement}
Data associated with this research are available and can be obtained by contacting the corresponding author. \\

\bibliographystyle{unsrt}  
\bibliography{references}

\end{document}